%% file: root.tex
\documentclass[letterpaper, 10pt, conference]{ieeeconf}  
\IEEEoverridecommandlockouts
\usepackage{cite}
\usepackage{graphicx}
\usepackage{array}
\usepackage{hyperref}
\usepackage{arydshln}
\usepackage{float}
\usepackage{blindtext}
\usepackage{hyphenat}
\usepackage{xspace}
\usepackage{lettrine}

\let\labelindent\relax

\input{cc_cdc-2023-generalized-plant.tex}

\title{Unified Behavioral Data-Driven Performance Analysis: A Generalized Plant Approach}

\author{L. M. Spin, C. Verhoek, W. P. M. H. Heemels, N. van de Wouw, R. T\'oth
\thanks{This research was supported by the Eindhoven Artificial Intelligence Systems Institute, Eindhoven University of Technology, the Netherlands and the European Union within the framework of the National Laboratory for Autonomous Systems (RRF-2.3.1-21-2022-00002). The research was also partially sponsored by the European Research Council under the Advanced ERC grant agreement PROACTHIS, no. 101055384.}%
\thanks{L. M. Spin, C. Verhoek and R. T\'oth are with the Control Systems Group, Eindhoven University of Technology, The Netherlands. R. T\'oth is also with the Institute for Computer Science and Control, Budapest, Hungary. W. P. M. H. Heemels is with the Control Systems Technology Group, Eindhoven University of Technology, The Netherlands. N. van de Wouw is with the Dynamics and Control Group, Eindhoven University of Technology, The Netherlands. Corresponding author: L. M. Spin, email address: \texttt{l.m.spin@tue.nl}.}}

\begin{document}
\maketitle

\begin{abstract}
    \input{section/abstract}
\end{abstract}
\pagenumbering{roman}
\section{Introduction}\label{sec:introduction}
\input{section/introduction}
\newpage\section{Problem setting}\label{sec:problemstatement}
\input{section/problemstatement}
\section{Data-driven generalized plant}\label{sec:mbanddd}
\input{section/mbanddd.tex}
\section{Dissipativity analysis}\label{sec:datadrivendissanalysis}
\input{section/datadrivendissanalysis}
\section{Examples}\label{sec:examples}
\input{section/examples}
\section{Conclusions}\label{sec:conclusion}
\input{section/conclusion}
\section*{Acknowledgement}
The authors would like to thank Julian Berberich for a fruitful discussion and his insightful comments regarding finite-horizon representations.
\appendix
\section{Appendix}\label{sec:appendix}
\input{section/appendix}

\bibliographystyle{IEEEtran}
\bibliography{ref_cdc-2023-generalized-plant}

\end{document}

%% file: cc_cdc-2023-generalized-plant.tex
\usepackage{tikz}
\usepackage{xspace}
\usepackage{xcolor}
\usepackage{amsthm}
\usepackage{amsmath,amssymb,amsfonts}
\usepackage{mathrsfs,mathtools}
\usepackage{mathalfa}
\usepackage{euscript}
\usepackage{csquotes}
\usepackage{enumitem}
\usepackage[normalem]{ulem}

\definecolor{mblue}{rgb}{0,0.4470,0.7410}
\definecolor{morange}{rgb}{0.8500,0.3250,0.0980}
\definecolor{myellow}{rgb}{0.9290,0.6940,0.1250}
\definecolor{mpurple}{rgb}{0.4940,0.1840,0.5560}
\definecolor{mgreen}{rgb}{0.4660,0.6740,0.1880}
\definecolor{mcyan}{rgb}{0.3010,0.7450,0.9330}
\definecolor{mred}{rgb}{0.6350,0.0780,0.1840}
\definecolor{mgreenblue}{rgb}{0.0,1.0,0.5}
\definecolor{parulablue}{rgb}{0.2431,0.1490,0.6588}
\definecolor{parulalblue}{RGB}{39,151,235}
\definecolor{parulagreen}{RGB}{129,204,89}
\definecolor{parulayellow}{RGB}{249,251,21}

\definecolor{cblue}{rgb}{0,0.9,1}
\definecolor{corange}{rgb}{1,0.7,0}

\theoremstyle{definition}
\newtheorem{defn}{Definition}
\newtheorem{exmp}{Example}
\theoremstyle{plain}
\newtheorem{theorem}{Theorem}
\newtheorem{lemma}{Lemma}

\newtheorem{assumption}{Assumption}
\theoremstyle{remark}

\newenvironment{definition}{\begin{defn}}{\hfill$\square$\end{defn}}

\newcommand{\comment}[1]{}

\newcounter{ass}

\newcommand{\mc}[1]{\mathcal{#1}}
\newcommand{\mf}[1]{\mathfrak{#1}}
\newcommand{\mr}[1]{\mathrm{#1}}
\newcommand{\mb}[1]{\mathbb{#1}}
\newcommand{\ms}[1]{\mathscr{#1}}

\DeclareFontFamily{U}{txcal}{\skewchar \font =45}
\DeclareFontShape{U}{txcal}{m}{n}{<-> txr-cal}{}
\DeclareMathAlphabet{\mathcalpxtx}{U}{txcal}{m}{n}

\newcommand{\posdef}{\succ}
\newcommand{\negdef}{\prec}
\newcommand{\possemidef}{\succcurlyeq}
\newcommand{\negsemidef}{\preccurlyeq}

\newcommand{\svdots}{\raisebox{0pt}{$\scalebox{.75}{\vdots}$}}
\newcommand{\sddots}{\raisebox{0pt}{$\scalebox{.75}{$\ddots$}$}}
\newcounter{Cx}
\newcommand{\itemC}{%
    \addtocounter{Cx}{1}
    \item[C\theCx)]}
\makeatletter
\newcommand*\bigcdot{\mathpalette\bigcdot@{1}}
\newcommand*\bigcdot@[2]{\mathbin{\vcenter{\hbox{\scalebox{#2}{$\m@th#1\bullet$}}}}}
\makeatother

\newcommand{\dnx}{n_\mr{x}}
\newcommand{\dny}{n_\mr{y}}
\newcommand{\dnu}{n_\mr{u}}

\newcommand{\Matrix}[3][1.25]{\ensuremath{
{\arraycolsep= #1 pt \def\arraystretch{1}
\left[ \begin{array}{#2} #3 \end{array} \right]}}}

%% file: section/abstract.tex
In this paper, we present a novel approach to combine data-driven non-parametric representations with model-based representations of dynamical systems. Based on a data-driven form of linear fractional transformations, we introduce a data-driven form of generalized plants. This form can be leveraged to accomplish performance characterizations, e.g., in the form of a mixed-sensitivity approach, and LMI-based conditions to verify finite-horizon dissipativity. In particular, we show how finite-horizon $\mc{\ell}_2$-gain under weighting filter-based general performance specifications are verified for implemented controllers on systems for which only input-output data is available. The overall effectiveness of the proposed method is demonstrated by simulation examples.

%% file: section/introduction.tex
\lettrine[findent=2pt]{\textbf{D}}{ }irect data-driven control is a generic term to categorize all control strategies that are based on measured data. The data is subsequently converted into control laws without system identification as an intermediate step. From a computational point of view, it is highly attractive to design control laws without identifying a mathematical model. Moreover, high-tech systems in, e.g., the semiconductor, aerospace, and process industry, are becoming increasingly more complex, causing first-principles modeling to become more and more challenging to accurately describe the behavior of these systems. This has motivated a trend towards data-driven control in recent years as an alternative for model-based approaches \cite{hou2013model}. 

A cornerstone result in direct data-driven analysis and control for discrete-time \textit{linear time-invariant} (LTI) systems is known as \textit{Willems' Fundamental Lemma} \cite{willems2005note}. This result relies on the behavioral system theory \cite{willems1997introduction} and characterizes the finite-horizon behavior of an LTI system using one \cite{berberich2020trajectory}, or multiple \cite{van2020willems}, measured input-output data trajectories. In order to systematically handle measurement noise on the data, quantified \cite{berberich2022quantitative} and robustified \cite{coulson2022robust} extensions of the Fundamental Lemma have been proposed. For a detailed overview of results based on Willems' Fundamental Lemma, we mention the survey \cite{markovsky2021behavioral}. These data-driven non-parametric representations of LTI systems have also been used to analyze system properties such as stability \cite{maupong2017lyapunov} and dissipativity \cite{romer2019one}, \cite{romer2017determining}. Tractable formulations for synthesizing controllers that are optimal with respect to various performance metrics have also been derived for, e.g., LQR and $\mc{H}_2$-control \cite{de2019formulas, van2020data, van2022quadratic, de2021low} and $\mc{H}_{\infty}$-control \cite{nouri2022discrete}. However, these approaches all assume direct state measurements, which are rarely available in practice. It must be noted that a remedy to this problem is suggested in \cite{de2019formulas}. However, as we will show in Section \ref{sec:problemstatement}, these results do not provide sufficient solutions for the dynamic output feedback problem. Recently, for input-output data, the feedback interconnection between a data-based plant and a model-based controller has been considered in \cite{wieler2021data}. In this work, the model-based controller is represented by a finite-horizon impulse response for which the analysis is limited to one degree of freedom SISO controllers without performance objectives incorporated in the design problem.

For model-based controller design, a systematic framework that allows the direct incorporation of performance specifications is known as the generalized plant framework \cite{zhou1998essentials}. Generalized plants contain all aspects of a control configuration except for the controller itself. In fact, the interconnection structure between the generalized plant and the controller is defined as a \emph{Linear Fractional Representation} (LFR) \cite{redheffer1960certain}. A powerful shaping tool for adding performance specifications to the generalized plant in the form of weighting filters is known as mixed-sensitivity shaping \cite{skogestad2005multivariable}. Appropriate choices of weighting filters can be used to characterize requirements related to bandwidth, phase-, gain-, and modulus margins in the frequency domain \cite{seron2012fundamental}. These frequency-domain concepts have an empirical relation to settling time and overshoot in the time domain \cite{horowitz2013synthesis}. This shaping approach has been successfully used to design optimal \cite{lewis2012optimal} and robust controllers \cite{green2012linear} in the industry. Moreover, the generalized plant framework allows for integral and high-frequency roll-off design by leveraging loop transformations \cite{bosgra2001design}. 

Such a systematic design approach with a priori performance guarantees is currently missing for the direct data-driven control framework. In this paper, we propose to unify data-driven representations of to-be-controlled systems based on Willems' Fundamental Lemma together with model-based representations of controllers and weighting filters with the purpose of analyzing the performance of the controllers via a mixed-sensitivity argument.

\newpage

The main contributions of this paper are as follows:
\begin{enumerate}
    \itemC Unify model-based and data-driven representations as one finite-horizon LFR-based representation,
    \itemC Derive tractable \emph{linear matrix inequality} (LMI)-based numerical methods to analyze general dissipativity-based performance of controllers,
    \itemC Demonstrate the strength of the new framework in two numerical case studies where the performance of given controllers is analyzed.
\end{enumerate}

The remainder of this paper is structured as follows. First, in Section~\ref{sec:problemstatement}, various system representations are introduced together with the considered problem statement. In Section~\ref{sec:mbanddd}, we show how model-based and data-based system representations can be unified as one representation. In Section~\ref{sec:datadrivendissanalysis}, we introduce dissipativity characterizations of the derived system representations for general performance analysis. We finalize the paper with examples in Section~\ref{sec:examples} followed by the conclusions in Section~\ref{sec:conclusion}.

\subsection{Notation}
The sets $\mathbb{R}$, $\mb{Z}$, and $\mathbb{N}$ denote the set of real numbers, integers, and non-negative integers, respectively. The open unit disc is denoted by $\mb{D}$, and its complement is denoted as $\mb{D}^\mr{c} := \mb{C} \setminus \mb{D}$. The zero matrix with dimension $n\times m$ and the square identity matrix with dimension $n$ are denoted by $\mathbf{0}_{n \times m}$ and $I_n$, respectively. For a  matrix $A \in \mathbb{R}^{n \times m}$ with $\mr{rank}(A)=r\leq \min \{n,m\}$, $A^\top \in \mathbb{R}^{m \times n}$ denotes its transpose and the nullspace is defined as $\mc{N}(A) := \{ x \in \mb{R}^{m} | A x = 0\}$. Furthermore, $A^{\perp}\in \mb{R}^{m \times m-r}$ denotes a matrix where the columns span $\mc{N}(A)$, i.e., $A A^{\perp} = 0_{n\times m-r}$. The set of symmetric matrices of dimension $n\times n$ is denoted by $\mb{S}^n$. The notation $A \posdef 0$, $A\possemidef 0$, $A \negdef 0$, and $A \negsemidef 0$ denotes that $A=A^\top$ is positive definite, nonnegative definite, negative definite and nonpositive definite, respectively. We denote $\mb{R}^{n \times m}[\xi]$ as the matrix-valued polynomial ring with $\xi$ the indeterminate. For brevity, we write quadratic products as $(*)^{\top} A x$, where $*$ denotes a symmetric term. Given a discrete-time signal $z \in (\mb{R}^{n_\mr{z}})^{\mb{Z}}$, where $(\mb{R}^{n_\mr{z}})^{\mb{Z}}$ defines the collection of all maps $z:\mb{Z}\rightarrow \mb{R}^{n_\mr{z}}$, the restriction to the interval $[0, L-1] \cap \mb{Z}$ is denoted by $z|_L$. The concatenation of two signals $z_1$ and $z_2$ is denoted by $z = z_1 \wedge z_2$. For a given sequence of samples $\{z_k\}_{k=0}^{N-1}$ such that $z(k) \in \mb{R}^{n_\mr{z}}$ for all $k \in [0, N-1]\cap \mb{N}$, the Hankel matrix $\mc{H}_L(z) \in \mb{R}^{L n_\mr{z} \times N-L+1}$ with depth $L$ is defined as
\begin{equation}
    \mc{H}_L(z) = \Matrix{cccc}{z(0) & z(1) & \hdots & z(N-L) \\
    z(1) & z(2) & \hdots & z(N-L+1)\\
    \vdots & \vdots & \ddots & \vdots\\
    z(L-1) & z(L) & \hdots & z(N-1)},
\end{equation}
for $0\leq L \leq N-1$. Lastly, $A \otimes B$ denotes the Kronecker product of two matrices $A$ and $B$.

%% file: section/problemstatement.tex
We consider a measured input-output trajectory $\mc{D}_N = \{u_k^d, y_k^d\}_{k=0}^{N-1}$ (data-dictionary) from some unknown LTI system $\mc{G}$, which is assumed to have a minimal state-space realization
\begin{equation}\label{eq:sys}
    \begin{aligned}
        (q x)(k) &= A x(k) + B u(k),\\
        y(k) &= C x(k) + D u(k).
    \end{aligned}
\end{equation}
Here $x(k)\in\mb{R}^{n_\mr{x}}, u(k)\in\mb{R}^{n_\mr{u}}$ and $y(k)\in\mb{R}^{n_\mr{y}}$ denote the state, input, and output, respectively, at time $k\in \mb{N}$. Moreover, $q$ denotes the forward shift operator, i.e., $(q x)(k) = x(k+1)$. The input-output behavior of $\mc{G}$, denoted by $\mf{B}$, is defined as the collections of all possible trajectories that are admissible solutions of the dynamics of $\mc{G}$:
\begin{multline}
    \mf{B} := \big\{(u,  y) \in (\mb{R}^{n_\mr{u}} \times \mb{R}^{n_\mr{y}})^{\mb{N}} \ \big| \\ \exists x \in (\mb{R}^{n_\mr{x}})^\mb{N} \text{ s.t. \eqref{eq:sys} holds for all } \ k \in \mb{N}\big\}.
\end{multline}
By eliminating the state of \eqref{eq:sys} using the Elimination Theorem \cite{willems1991paradigms}, the behavior is equivalently characterized by a time-domain operator $P(q)$ such that
\begin{equation}
    \mf{B} = \left\{(u,  y) \in (\mb{R}^{n_\mr{u}} \times \mb{R}^{n_\mr{y}})^{\mb{N}} \ \bigg| \ P(q) \Matrix{c}{u \\ y} = 0\right\},
    \label{eq:kernel}
\end{equation}
which is referred to as the kernel representation. Note that $P(\xi)\in\mb{R}^{n_\mr{y}\times n_\mr{u}+n_\mr{y}}[\xi]$, with indeterminate $\xi$, is a matrix-valued polynomial function, which is defined as $P(\xi) = \sum_{i=0}^{\ell}P_i \xi^i$ where $\ell$ is usually referred to as the lag of the system. When $\xi$ is taken as $q$ and $\ell$ is minimal among all possible kernel representations that characterize $\mf{B}$, then $P(q)$ denotes a minimal kernel representation of $\mc{G}$. Specifically, the lag is defined as the observability index \cite{willems1991paradigms}.

After presenting model-based representations of LTI systems, we present a data-driven non-parametric representation of LTI systems. In order to provide a data-driven representation of LTI systems we require a notion of persistence of excitation for signals.
\begin{definition}[Persistence of Excitation]
    A sequence $\{u_k\}_{k=0}^{N-1}$ where $u(k)\in \mb{R}^{n_\mr{u}}$ for all $k \in [0, \cdots, N-1]$ is said to be persistently exciting of order $L$ if $\mc{H}_L(u)$ has full row rank.
\end{definition}
Under a persistence of excitation condition on the input of $\mc{D}_N$, we can represent LTI systems using \emph{only} the data-set $\mc{D}_N$ \cite{willems2005note}, which is formalized for controllable LTI systems of the form~\eqref{eq:sys} in \cite{berberich2020trajectory} by the following lemma.
\begin{lemma}[Fundamental Lemma, \cite{berberich2020trajectory}]
Given a measured input-output trajectory $\mc{D}_N = \{u_k^d, y_k^d\}_{k=0}^{N-1}$ of a controllable LTI system $\mc{G}$ for which the input sequence $\{u^d_k\}_{k=0}^{N-1}$ is persistently exciting of order $L+n_\mr{x}$, the sequence $\{u_k, y_k\}_{k=0}^{L-1}$ is a trajectory of $\mc{G}$ if and only if there exists $g \in \mb{R}^{N-L+1}$ such that 
\begin{equation}
    \Matrix{c}{\mc{H}_L(u^d) \\ \mc{H}_L(y^d)} g = \Matrix{c}{u|_L \\ y|_L}.
    \label{eq:fundLemRepre}
\end{equation}
\end{lemma}
\noindent
This result has been generalized in \cite{markovsky2022identifiability} for which controllability or strict assumptions on the persistence of excitation of the input are not required. It is shown that the finite-horizon behavior
\begin{multline}
    \mf{B}|_L := \{ (u_p, y_p) \in (\mb{R}^{n_\mr{u}} \times \mb{R}^{n_\mr{y}})^L \ | \ \exists (u_f, y_f) \in \\ (\mb{R}^{n_\mr{u}} \times \mb{R}^{n_\mr{y}})^{\mb{N}} \text{ s.t. }  (u_p, y_p)\wedge (u_f, y_f) \in \mf{B} \},
\end{multline}
is equal to the image of $\Matrix{c}{\mc{H}_L(u^d) \\ \mc{H}_L(y^d)}$ if and only if
\begin{equation}
\mr{rank} \left( \Matrix{c}{\mc{H}_L(u^d) \\ \mc{H}_L(y^d)}\right) = n_\mr{u} L + n_\mr{x}, \label{eq:fundlemrank}
\end{equation}
is satisfied for $L\geq \ell$. If this rank condition is satisfied, the Hankel matrices provide a data-driven non-parametric representation of the system. If input-state measurements are available instead of input-output measurements, the rank condition of \eqref{eq:fundlemrank} is relaxed to only require a rank of $n_\mr{x}+n_\mr{u}$. This condition has been used extensively in \cite{de2019formulas} to obtain LMI-based conditions to verify stability and synthesize controllers on the basis of data. In \cite{de2019formulas}, it was suggested that, on the basis of input-output data, an extended state can be constructed as follows:
\begin{equation}
    \chi_L(k) = \mr{col}\left( u_{[k-L, k-1]}, y_{[k-L, k-1]} \right),
\end{equation}
such that the extended state-space representation of system \eqref{eq:sys} is defined as
\begin{equation}
\begin{aligned}
    \chi_L(k+1) &= \ms{A} \chi_L(k) + \ms{B} u(k),\\
    y(k) &= \ms{C} \chi_L(k) + \ms{D} u(k).
\end{aligned}\label{eq:extendsys}
\end{equation}
If the initial condition $\chi_0$ of the extended state-space representation is an element of $\mf{B}|_L$ and it holds that $L \geq \ell$, then the finite-horizon behavior of \eqref{eq:extendsys} is equal to the finite-horizon behavior of the original system \eqref{eq:sys}. The advantage of this representation is that the state can be constructed directly from the input-output measurements. It is argued that if the rank condition
\begin{equation}
    \mr{rank} \left( \Matrix{c}{\mc{H}_1(u^d) \\ \mc{H}_1(\chi_L^d)}\right) = L(n_\mr{u}+n_\mr{y}) + n_\mr{u},
    \label{eq:ExtendRank}
\end{equation}
is satisfied, the input-state results provided in \cite{de2019formulas} are directly applicable for the extended state-space representation. Subsequently, the synthesized state-feedback controller for the extended state-space formulation is then realized as a dynamic output-feedback controller for the original system. Despite the simplicity of this concept, Rank Condition \eqref{eq:ExtendRank} does not hold in general, since the extended state-space representation is not controllable in general. We show this by means of a randomly generated example system with $(\dnu,\dnx,\dny)=(2,20,3)$ and $\ell=7$.
\begin{figure}[t]
\centering
\includegraphics[width=\linewidth]{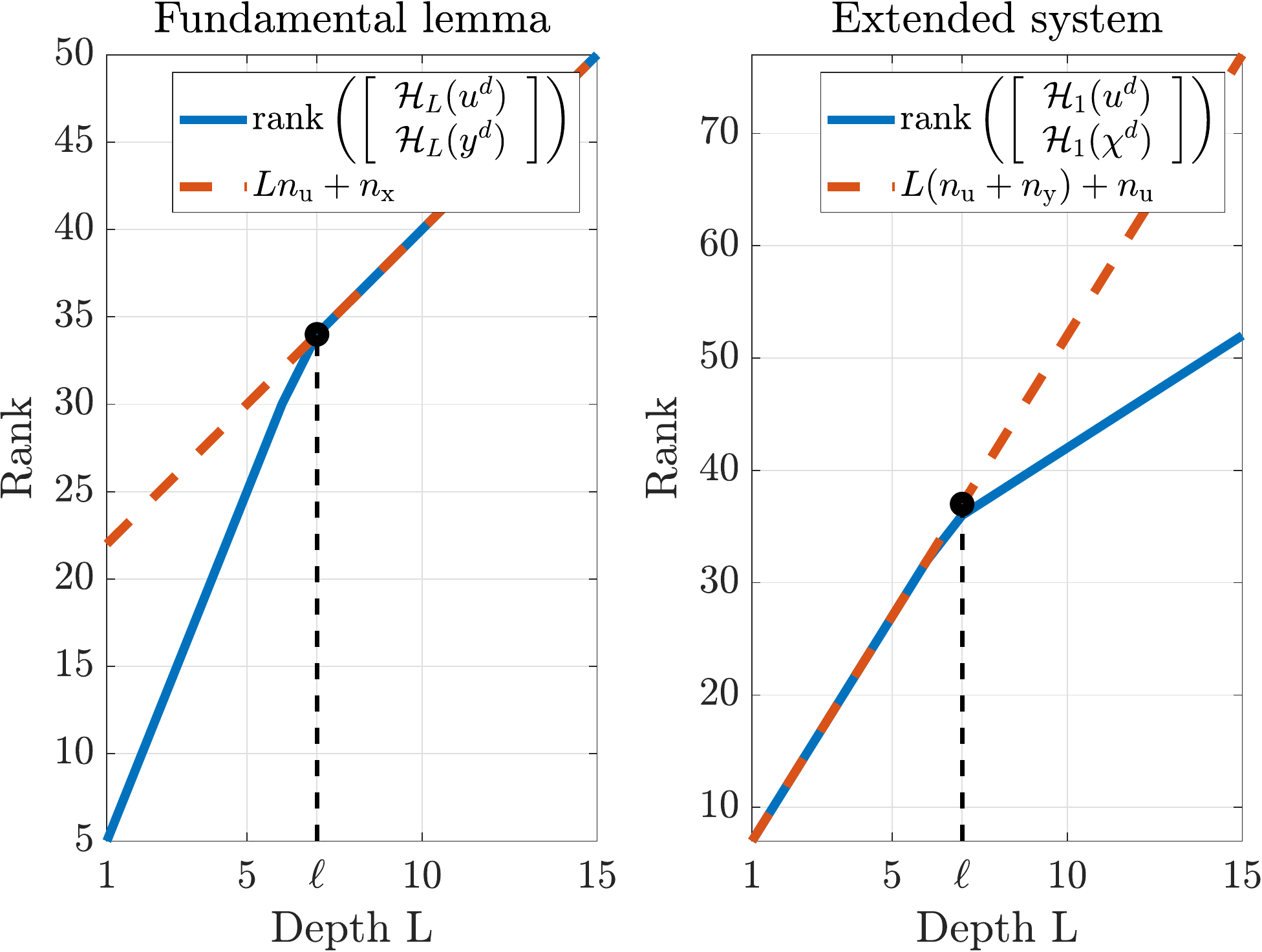}
\caption{Rank condition on data vs mathematical rank condition.} 
\label{fig:FundVSExtend}
\end{figure}
In Fig.~\ref{fig:FundVSExtend}, it is shown that the Rank Condition \eqref{eq:fundlemrank} for this system is satisfied for a depth $L$ larger than or equal to the lag of the system, while the Rank Condition \eqref{eq:ExtendRank} based on the extended state realization of this example system is only satisfied for a depth $L$ smaller than or equal to the lag of the system. By means of this simple example, it is shown that the Rank Condition \eqref{eq:ExtendRank} for the extended state-space representation can generally not be satisfied for depth $L$ larger than or equal to the lag, while a larger depth is required by the Fundamental Lemma in order for the data to represent the system. More importantly, in practice the lag is generally unknown, hence an upper bound has to be chosen. We conclude that LMI-based stability and synthesis results derived in \cite{de2019formulas}, in their current form, are not sufficient to address the output-feedback control synthesis problem for general LTI systems. Therefore, there is a need for direct data-driven performance analysis methods, which consider input-output data.

\subsection{Problem statement}
Consider a controllable system with behavior $\mf{B}$ from which we measured the data-dictionary $\mc{D}_N$ that satisfies~\eqref{eq:fundlemrank}. In this work, we aim to explore an alternative way to use the Fundamental Lemma directly to solve the problem of LMI-based dissipativity analysis of the interconnection of model-based representations, e.g., controllers and weighting filters, and the data-driven representation of $\mf{B}$.

%% file: section/mbanddd.tex
In this section, we provide a representation of general interconnections between data-driven and model-based representations. Specifically, we combine model-based input-output representations with data-driven non-parametric representations provided by the Fundamental Lemma. First, input-output representations are formulated for general control interconnections, which are subsequently interconnected with data-driven representations.

\subsection{Model-based representations}

\begin{figure}[t]
\centering
\includegraphics[width=.7\linewidth]{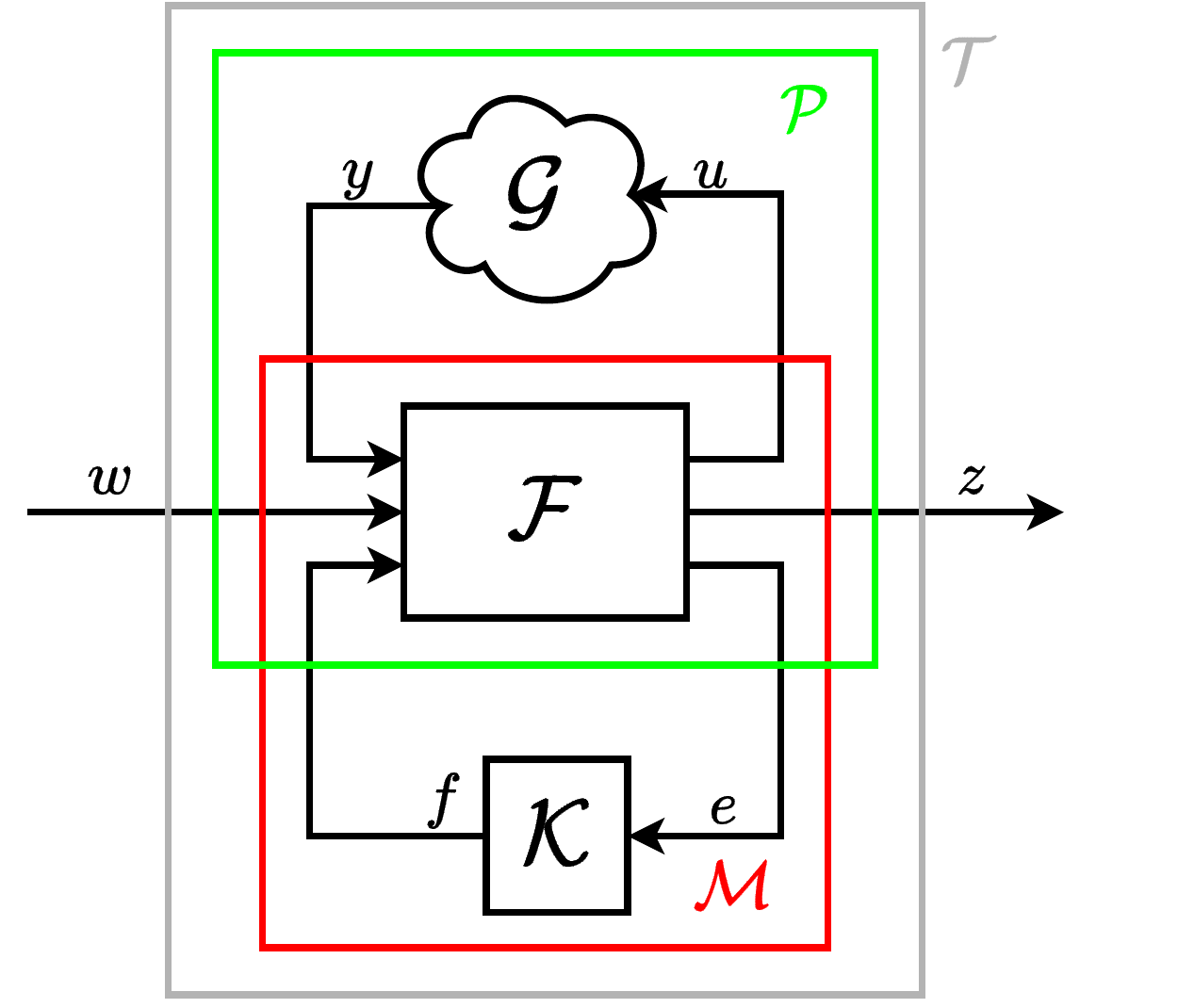}
\caption{Linear fractional representation of general interconnections between data-driven and model-based representations.} 
\label{fig:DDMBGenPlant}
\end{figure}

The polynomial matrix $P(\xi)$ of the kernel representation~\eqref{eq:kernel} is partitioned as $P(\xi) = \Matrix{cc}{-N(\xi) \ & \ D(\xi)}$, such that $D(q) y = N(q) u$ defines an input-output representation. In order to argue about minimality and characterize system properties of input-output representations, we give the following definition of left primeness.
\begin{definition}[Left prime \cite{willems2007behaviors}]
    A polynomial matrix $P(\xi) \in \mb{R}^{n \times m}[\xi]$ is said to be left prime over $\mb{R}[\xi]$ if and only if rank$(P(\lambda)) = n$ for all $\lambda \in \mb{C}$.
\end{definition}
Two polynomial matrices $D(\xi)$ and $N(\xi)$, are said to be coprime over $\mb{R}[\xi]$ if the polynomial matrix $[-N(\xi) \ D(\xi)]$, is left prime over $\mb{R}[\xi]$. For the well-known system theoretic properties of asymptotic stability, controllability, and stabilizability \cite{willems1991paradigms} we provide verifiable conditions in the following theorem.
\begin{theorem}[System properties \cite{willems1991paradigms}]
    Consider a system $\mc{G}$ with input-output (IO) representation $D(q) y = N(q) u$, system theoretic properties can be verified by the following conditions: 
    \begin{enumerate}
        \item \textbf{Asymptotically stable}, if and only if each root of $\mr{det}(D(\xi))$ resides within $\mb{D}$,
        \item \textbf{Controllable}, if and only if $D(\xi)$ and $N(\xi)$ are coprime,
        \item \textbf{Stabilizable}, if and only if rank$([-N(\lambda) \ D(\lambda)])$ is the same for all $\lambda \in \mb{D}^\mr{c}$.
    \end{enumerate}
\end{theorem}
We consider general control interconnections between model-based and data-driven representations. Such interconnections are defined as \emph{Linear Fractional Representations} (LFR) and are shown in Fig.~\ref{fig:DDMBGenPlant}, where the rectangular blocks denote model-based input-output representations and the cloud-shaped blocks denote data-based representations. 
In this LFR structure, all data-based components are extracted into the upper block $\mc{G}$ and all model-based controller components into the lower block $\mc{K}$. The remaining block $\mc{F}$ specifies the control configuration and contains additional weighting filters. The control configuration and controller blocks have the following model-based input-output representation
\begin{equation}
\begin{aligned}
\mc{F}&: \left\{\begin{aligned}
D_\mr{u}^\mr{f}(q) u &= N_\mr{u y}^\mr{f}(q) y + N_\mr{u w}^\mr{f} (q) w + N_\mr{u f}^\mr{f}(q) f,\\
D_\mr{z}^\mr{f}(q) z & = N_\mr{z y}^\mr{f}(q) y + N_\mr{z w}^\mr{f} (q) w + N_\mr{z f}^\mr{f}(q) f,\\
D_\mr{e}^\mr{f}(q) e & = N_\mr{e y}^\mr{f}(q) y + N_\mr{e w}^\mr{f} (q) w + N_\mr{e f}^\mr{f}(q) f,
\end{aligned}\right.\\
 \mc{K}&: \big\{ \ D_\mr{f}^\mr{k}(q) f = N_\mr{f e}^\mr{k}(q)e.
 \end{aligned}
\end{equation}
The following assumption on $\mc{F}$ and $\mc{K}$ is used such that the interconnection between them can be specified.
\begin{assumption}
    Is it assumed that the pairs $(D_{\bigcdot}^\mr{f}, N_{\bigcdot y}^\mr{f})$, $(D_{\bigcdot}^\mr{f}, N_{\bigcdot w}^\mr{f})$, $(D_{\bigcdot}^\mr{f}, N_{\bigcdot f}^\mr{f})$, and $(D_\mr{f}^\mr{k}, N_\mr{f e}^\mr{k})$ are coprime and that the determinant of the denominator polynomials matrices $D_\mr{u}^\mr{f}(\xi), D_\mr{z}^\mr{f}(\xi), D_\mr{e}^\mr{f}(\xi)$, and $D_\mr{f}^\mr{k}(\xi)$ are not the zero polynomial.
    \label{assump:coprimeLFR}
\end{assumption}
Define $D(q) = \mr{diag}(D_\mr{u}^\mr{f}(q), D_\mr{z}^\mr{f}(q), D_\mr{e}^\mr{f}(q))$ and
\begin{equation}
    N(q) = \Matrix{ccc}{N_\mr{u y}^\mr{f}(q) & N_\mr{u w}^\mr{f} (q) & N_\mr{u f}^\mr{f}(q) \\
    N_\mr{z y}\mr{f}(q) & N_\mr{z w}^\mr{f} (q) & N_\mr{z f}^\mr{f}(q) \\
    N_\mr{e y}^\mr{f}(q) & N_\mr{e w}^\mr{f} (q) & N_\mr{e f}^\mr{f}(q)},
\end{equation}
and note that $(D(q), N(q))$ is not guaranteed to be controllable, but by Assumption \ref{assump:coprimeLFR} it is stabilizable. The interconnection between $\mc{F}$ and $\mc{K}$ is defined through a \emph{Linear Fractional Transformation} (LFT). Usually, LFTs are defined for state-space or transfer function representations, we define the LFT for input-output representations.
\begin{lemma}[LFT \cite{vidyasagar2011control}]
The lower LFT 
\begin{equation}
\hspace{-15pt}\resizebox{.93\hsize}{!}{$
     \mc{M} = \ms{F}_\ell(\mc{F}, \mc{K}): \left\{\begin{aligned}
D_\mr{u}^\mr{m}(q) u &= N_\mr{u y}^\mr{m}(q) y + N_\mr{u w}^\mr{m}(q) w,\\
D_\mr{z}^\mr{m}(q) z &= N_\mr{z y}^\mr{m}(q) y + N_\mr{z w}^\mr{m}(q) w,
\end{aligned}\right.
\label{eq:MBM}$}
\end{equation}
exists and is well-posed if there exists a unimodular $U \in \mb{R}^{n_\mr{f} \times n_\mr{f}}[\xi]$ such that
\begin{equation}
    \hspace{-2pt}Q(\xi) = U(\xi) \left( D_\mr{f}^\mr{f}(\xi) - N_\mr{f e}^\mr{f}(\xi) \left( D_\mr{e}^\mr{k}(\xi) \right)^{-1} N_\mr{e f}^\mr{k}(\xi) \right)^{-1}
\end{equation}
is unimodular and $U(\xi)N_\mr{f e}^\mr{f}(\xi) \left( D_\mr{e}^\mr{k}(\xi) \right)^{-1} \in \mb{R}^{n_\mr{f}\times n_\mr{e}}[\xi]$.
\end{lemma}
\begin{proof}
    The rational matrix function $\left( D_\mr{e}^\mr{f}(\xi) \right)^{-1} $ exists since $\mr{det}(D_\mr{e}^\mr{f}(\xi)) \ne 0$ over $\mb{R}[\xi]$ by Assumption \ref{assump:coprimeLFR}. Rewriting the thrid equation of $\mc{F}$ in rational form and subsequently substituting it into $\mc{K}$ gives that
    \begin{multline*}
        (D_\mr{f}^\mr{f}(q) - N_\mr{f e}^\mr{f}(q) \left( D_\mr{e}^\mr{k}(q) \right)^{-1} N_\mr{e f}^\mr{k}(q)) f = \\ N_\mr{f e}^\mr{f}(q) \left( D_\mr{e}^\mr{k}(q) \right)^{-1} \left(N_\mr{f}^\mr{f}(q) y + N_\mr{e w}^\mr{f} (q) w\right),
    \end{multline*}
    which is a well-defined input-output representation if pre-multiplied with a unimodular matrix $U(\xi)$ such that $Q(\xi)$ is a unimodular matrix and $U(\xi)N_\mr{f e}^\mr{f}(\xi) \left( D_\mr{e}^\mr{k}(\xi) \right)^{-1}$ is a polynomial matrix since premultiplication with unimodular matrices does not change the behavior \cite{willems1997introduction}. The inverse of a unimodular matrix is a polynomial matrix and hence
    \begin{equation*}
        e = Q^{-1}(q) U(q) \left( N_\mr{e y}^\mr{f}(q) y + N_\mr{e w}^\mr{f} (q) w \right),
    \end{equation*}
    defines a well-posed interconnection that can be substituted in the remaining two equations of $\mc{F}$ which provides a well-defined input-output representation of $\mc{M}$.
\end{proof}

\subsection{Generalized plants with finite-horizon representations}\label{sec:DDGP}
We now derive a finite-horizon \emph{upper} LFT between the data-based representation $\mc{G}$ and model-based representation $\mc{F}$. In order to define the interconnection with model-based and data-based representations, the length of signal trajectories needs to align between the different representations. For a polynomial matrix $P(\xi) = \sum_{i=0}^\ell P_i \xi^i \in\mb{R}^{n\times m}[\xi]$, define the following upper-Toeplitz matrix $\ms{T}_L(P) \in \mb{R}^{(L-\ell)n\times L m}$ as
\begin{equation}
    \ms{T}_L(P) = \Matrix{ccccccc}{P_0 & P_1 & \cdots & P_\ell & 0 & \cdots & 0 \\ 0 & P_0 & P_1 & \cdots & P_\ell & \ddots & \vdots \\ \vdots & \ddots & \ddots & \ddots & & \ddots & 0 \\ 0 & \cdots & 0 & P_0 & P_1 & \cdots & P_\ell},
    \label{eq:polyToep}
\end{equation}
such that $\ms{T}_L(P)$ contains $L$ block-columns. This matrix has been used in \cite{markovsky2010closed} to constrain a model-based representation to finite-horizon trajectories such that for an input-output representation, it holds that
\begin{equation}
    \ms{T}_L(D) y|_L = \ms{T}_L(N) u|_L,
\end{equation}
for all $L \geq \ell + 1$. For brevity of notation, we denote $\ms{T}_L(D_{\bullet}^\mr{f})$ and $\ms{T}_L(N_{\bullet}^\mr{f})$ by $T_{\bullet}^\mr{D}$ and $T_{\bullet}^\mr{N}$, respectively, in the following lemma.
\begin{lemma}[Generalized Plant]\label{lem:Repr}
The finite-horizon plant $\mc{P}|_L$, defined by the upper LFT $\ms{F}_u(\mc{G}|_L, \mc{F}|_L)$ such that for any $w|_L \in \mb{R}^{L n_\mr{w}}$ and $z|_L \in \mb{R}^{L n_\mr{z}}$ there exists a $g\in\mb{R}^{N-L+1}$ such that
\begin{equation}
\begin{aligned}
T_\mr{u}^\mr{D} \mc{H}_L(u^\mr{d}) g &= T_\mr{u y}^\mr{N} \mc{H}_L(y^\mr{d}) g + T_\mr{u w}^\mr{N} w|_L + T_\mr{u f}^\mr{N} f|_L,\\
T_\mr{z}^\mr{D} z|_L & = T_\mr{z y}^\mr{N} \mc{H}_L(y^\mr{d}) g + T_\mr{z w}^\mr{N} w|_L + T_\mr{z f}^\mr{N} f|_L,\\
T_\mr{e}^\mr{D} e|_L & = T_\mr{e y}^\mr{N} \mc{H}_L(y^\mr{d}) g + T_\mr{e w}^\mr{N} w|_L + T_\mr{e f}^\mr{N} f|_L,
\end{aligned}\label{eq:FinHorizonRepr}
\end{equation}
is called a generalized plant, if there exists a controller $\mc{K}$ such that the closed-loop interconnection $\mc{T}|_L := \ms{F}_\ell(\mc{P}|_L, \mc{K}|_L)$ is stable.
\end{lemma}
\begin{proof}
Consider the upper-Toeplitz representation of $\mc{F}$ and substitute $u|_L = \mc{H}(u^\mr{d}) g$ and $y|_L = \mc{H}(y^\mr{d}) g$, which provides the finite-horizon representation $\mc{P}|_L$ given in \eqref{eq:FinHorizonRepr}. The solution space for $g$ is defined as the quotient space given by the particular solution $g_0$, e.g., the minimal norm solution, and a subspace equal to the nullspace of $\Matrix{cc}{\mc{H}_L(u^\mr{d})^{\top} & \mc{H}_L(y^\mr{d})^{\top}}^{\top}$. This quotient space together with admissible finite-horizon trajectories of $f|_L, w|_L, z|_L$, and $e|_L$ exactly parametrize the finite-horizon behavior $\mf{B}|_L$ associated to $\mc{P}|_L$ which is shown in \cite{markovsky2010closed}. If there exists a controller $\mc{K}$ such that the closed-loop system is stable, $\mc{P}|_L$ is called a generalized.
\end{proof}

This result allows for a general LFR-based interconnection framework between data-driven and model-based representations. In the following section, we aim at deriving tractable conditions to verify dissipativity of such representations.

%% file: section/datadrivendissanalysis.tex

\subsection{Data-driven dissipativity analysis}
The dissipativity notion introduced in \cite{hill1980dissipative} is defined for input-output representations of dynamical systems. In particular, this notion does not require a storage function and is equivalent to the notion of dissipativity of \cite{willems1972dissipative1}, \cite{willems1972dissipative2}.
\begin{definition}[Dissipativity, \cite{hill1980dissipative}]
A system $\mc{G}$, given by \eqref{eq:sys}, with input $u:\mb{N} \rightarrow \mb{R}^{n_\mr{u}}$, output $y:\mb{N} \rightarrow \mb{R}^{n_\mr{y}}$ and behavior $\mf{B}$ is dissipative with respect to the supply rate $\Pi \in \mb{S}^{(n_\mr{u}+n_\mr{y})\times (n_\mr{u}+n_\mr{y})}$, if and only if it holds that
\begin{equation}
    \sum_{k=0}^{r}(*)^{\top}\Pi\Matrix{c}{u(k) \\ y(k)} \geq 0, \quad \text{for all} \  r \in\mb{N},
\end{equation}
for all trajectories $(u, y) \in \mf{B}$ with initial condition $x_0=0$.
\end{definition}
The supply rate, defining a quadratic supply function, is usually partitioned as
\begin{equation}
    \Pi:=\Matrix{cc}{Q & S  \\ S^{\top} & R},
\end{equation}
with $Q \in \mb{S}^{n_\mr{u}}, R\in \mb{S}^{n_\mr{y}}$, and $S \in \mb{R}^{n_\mr{u}\times n_\mr{y}}$. Dissipativity generalizes a large class of analysis metrics such as passivity properties for $(Q,S,R) = (0, I, 0)$, input feedforward passivity for $(Q,S,R) = (-v I, I, 0)$, passivity shortage for $(Q,S,R) = (0, I, \beta I)$, and the upperbound $\gamma$ on the $\ell_2$-gain with $(Q,S,R) = (\gamma^2 I, 0, -I)$ of the system.

In \cite{maupong2017lyapunov}, the classical dissipativity notion has been reformulated for finite-horizon trajectories, which aligns better with measured trajectories of a system with a finite number of samples.
\begin{definition}[$L$-Dissipativity \cite{maupong2017lyapunov}]
The system $\mc{G}$ is said to be $L$-dissipative with respect to the supply rate $\Pi$ if
\begin{equation}
    \sum_{k=0}^{L-1}(*)^{\top}\Pi\Matrix{c}{u(k) \\ y(k)} \geq 0,
    \label{eq:LdissSum}
\end{equation}
for all trajectories $\{ u_k, y_k \}_{k=0}^{L-1}$ of $\mc{G}$ with initial condition $x_0 = 0$.
\label{def:Ldiss}
\end{definition}
It has been shown in \cite{koch2021determining} that, under mild conditions, by taking the limit $L\rightarrow \infty$, $L$-dissipativity is equivalent to classical dissipativity. The notion of finite-horizon dissipativity in Definition \ref{def:Ldiss}, enables the use of the Fundamental Lemma for dissipativity analysis in a straightforward manner. Rewriting the sum \eqref{eq:LdissSum} using the Kronecker product provides the following compact notation for $L$-dissipativity:
\begin{equation}
{\small    
(*)^{\top} \Pi_L\Matrix{c}{u|_L \\ y|_L} = (*)^{\top} \Matrix{cc}{I_L \otimes Q & I_L \otimes S \\ I_L \otimes S^{\top} & I_L \otimes R}\Matrix{c}{u|_L \\ y|_L}\geq 0}.
    \label{eq:krondiss}
\end{equation}
We know that any length-$L$ trajectory of $\mc{G}$ is a linear combination of the columns of the Hankel matrices in \eqref{eq:fundLemRepre}. Moreover, as in \cite{romer2019one}, we can enforce the zero initial condition of the state of some minimal state-space realization of $\mc{G}$ by only considering trajectories $\mc{G}$ with length $L$ in which the first $\nu\geq \ell(\mc{G})$ samples are all zero. The space of all trajectories of $\mc{G}$ with length $L$ and the first $\nu$ samples zero is given by
\begin{equation}
    V_L^\nu(u,y) \Matrix{c}{\mc{H}_L(u^d) \\ \mc{H}_L(y^d)} g = 0,
    \label{eq:zeroiniconstr}
\end{equation}
where
\begin{equation}
\begin{aligned}
    V_L^\nu(u,y) &= \Matrix{cc}{V_L^\nu(u) & 0_{\nu n_\mr{u} \times L n_\mr{y}} \\ 0_{\nu n_\mr{y} \times L n_\mr{u}} & V_L^\nu(y)},\\ V_L^\nu(u) &= \Matrix{cc}{I_{\nu n_\mr{u}} & 0_{\nu n_\mr{u} \times n_\mr{u}(L-\nu)}},\\ V_L^\nu(y) &= \Matrix{cc}{I_{\nu n_\mr{y}} & 0_{\nu n_\mr{y} \times n_\mr{y}(L-\nu)}}.
\end{aligned}
\end{equation}
In order to reformulate the dissipativity inequality \eqref{eq:krondiss} with the zero initial conditions constraint \eqref{eq:zeroiniconstr} into a tractable condition, we introduce Finsler's Lemma.
\begin{lemma}[Finsler's Lemma \cite{de2007stability}]
Let $x \in \mb{R}^n, Q \in \mb{S}^n$ and $B \in \mb{R}^{m \times n}$ such that rank$(B) < n$. The following statements are equivalent:
\begin{itemize}
    \item $x^{\top} Q x \geq 0$ for all $x$ satisfying $B x = 0$.
    \item $(B^{\perp})^{\top} Q B^{\perp} \succeq 0$.
    \item There exists a $\mu \in \mb{R}$ such that $Q+ \mu  B^{\top} B \succeq 0$.
    \item There exists an $X \in \mb{R}^{n\times m}$ such that $X B+ B^{\top} X^{\top} - Q \preceq 0$.
\end{itemize}
\end{lemma}
Using Finsler's Lemma, inequality \eqref{eq:krondiss} and constraint \eqref{eq:zeroiniconstr} can be written as the LMI-condition 
\begin{equation}
    (*){^\top} \Pi_L \Matrix{c}{\mc{H}_L(u^d) \\ \mc{H}_L(y^d)} \left( V_L^\nu \Matrix{c}{\mc{H}_L(u^d) \\ \mc{H}_L(y^d)} \right)^{\perp} \succeq 0 ,\label{eq:DissLMI}
\end{equation}
which can be verified numerically. The data-driven LMI characterization of dissipativity \eqref{eq:DissLMI} will be combined with model-based representations to analyze the time-domain performance of controllers in the following section.

\subsection{Dissipativity analysis for combined generalized plants}
In this paper, we assume that the controller $\mc{K}$ is given for a certain plant $\mc{G}$, for which only measurements are available. Hence, we are interested in an LMI-based dissipativity analysis method of the interconnection between a model-based representation of $\mc{M}$ given in \eqref{eq:MBM} and a data-driven representation of $\mc{G}$ given in \eqref{eq:fundLemRepre}. The finite-horizon representation of $\mc{M}$ is defined as
\begin{equation}
\mc{M}|_L: \left\{\begin{aligned}
T_\mr{u}^{\mr{D}} u|_L &= T_\mr{u y}^{\mr{N}} y|_L + T_\mr{u w}^{\mr{N}} w|_L,\\
T_\mr{z}^{\mr{D}} z|_L & = T_\mr{z y}^{\mr{N}} y|_L + T_\mr{z w}^{\mr{N}} w|_L,
\end{aligned}\right.\label{eq:FiniteM}
\end{equation}
where we denote $\ms{T}_L(D^\mr{m}_{\bullet})$ and $\ms{T}_L(N^\mr{m}_{\bullet})$ by $T^{\mr{D}}_{\bullet}$ and $T^{\mr{N}}_{\bullet}$, respectively. The following lemma provides a constraint that guarantees that $\mc{M}|_L$ is initially at rest, which is required for the subsequent dissipativity analysis.
\begin{lemma}\label{lem:DDMBIni}
    If the constraint $V_L^\nu(w,z)\Matrix{c}{w\\z} = 0$ is imposed, $\mc{M}|_L$ is initially at rest if and only if $\nu > \ell(\mc{M})$.
    \label{lem:init}
\end{lemma}
\begin{proof}
By enforcing this constraint, the first $\nu$ samples of the signals $w$ and $z$ are identically zero, based on the definition of the upper-Toeplitz matrices \eqref{eq:polyToep}. Now consider the equation specified by the second row in \eqref{eq:FiniteM}, i.e., 
\begin{equation*}
    T_\mr{z}^{\mr{D}}z|_L = T_\mr{z y}^{\mr{N}} y|_L + T_\mr{z w}^{\mr{N}} w|_L.
\end{equation*}
The first block row of the matrices $T_\mr{z}^{\mr{D}},  T_\mr{z y}^{\mr{N}},$ and $ T_\mr{z w}^{\mr{N}}$ contain at most $\ell(\mc{M})$ non-zero blocks by definition. Since $\nu > \ell(\mc{M})$, it follows that $T_\mr{z}^{\mr{D}} z|_L=  T_\mr{z w}^{\mr{N}} w|_L =0 $. We conclude that $T_\mr{z y}^{\mr{N}} y|_L = 0$ must hold as well, which implies that the first $\ell(\mc{M})$ samples of $y|_L$ are zero. The derivation that the first $\ell(\mc{M})$ samples of $u|_L$ are zero also follows along the same lines for the first block row of \eqref{eq:FiniteM}.
\end{proof}
By Lemma \ref{lem:Repr}, the finite-horizon closed-loop system $\mc{T}|_L = \ms{F}_\mr{u} (\mc{G}|_L, \mc{M}|_L)$ is defined as
\begin{equation}
\mc{T}|_L: \left\{\begin{aligned}
T_\mr{u}^{\mr{D}} \mc{H}_L(u^\mr{d}) g &= T_\mr{u y}^{\mr{N}} \mc{H}_L(y^\mr{d}) g + T_\mr{u w}^{\mr{N}} w|_L,\\
T_\mr{z}^{\mr{D}} z|_L & = T_\mr{z y}^{\mr{N}} \mc{H}_L(y^\mr{d}) g + T_\mr{z w}^{\mr{N}} w|_L.
\end{aligned}\right.\label{eq:FiniteT}
\end{equation}
Note that this mixed representation does not require us to specify what $w|_L$ and $z|_L$ are, and hence we are able to characterize dissipativity of the closed-loop system for arbitrary input and output pairs. The following theorem gives an LMI condition to verify dissipativity of the interconnection between model-based and data-driven representations as an LFR.
\begin{theorem}
The closed-loop finite-horizon system $\mc{T}|_L = \ms{F}_\mr{u}(\mc{G}|_L, \mc{M}|_L)$ is $L$-dissipative with respect to the supply rate $\Pi$, if
\begin{equation}\label{eq:DDMBLMI}
    (*)^{\top} \Matrix{ccc}{0 & 0 & \\ 0 & \Pi_L} B^{\perp} \succeq 0,
\end{equation}
where
\begin{equation}
    B = \Matrix{ccc}{T_\mr{uy}^{\mr{N}} \mc{H}_L(y^d)  -T_\mr{u}^{\mr{D}} \mc{H}_L(u^d) & T_\mr{uw}^{\mr{N}} & 0 \\ T_\mr{z y}^{\mr{N}} \mc{H}_L(y^d) & T_\mr{z w}^{\mr{N}} & -T_\mr{z}^{\mr{D}} \\ 0 & V_L^\nu(w) & 0 \\ 0 & 0 & V_L^\nu(z)}.
    \label{eq:B}
\end{equation}
\label{thm:DDMBLMI}
\end{theorem}
\begin{proof}
    Note that
    \begin{equation}
        (*)^{\top} \Matrix{cc}{0 & 0 \\ 0 & \Pi_L} \Matrix{c}{g \\ w|_L \\ z|_L}  \geq 0
    \end{equation}
    is equivalent to \eqref{eq:krondiss}. By rewriting the finite-horizon representation \eqref{eq:FiniteT}, we have that
    \begin{equation}
        \Matrix{ccc}{T_\mr{uy}^{\mr{N}} \mc{H}_L(y^d)  -T_\mr{u}^{\mr{D}} \mc{H}_L(u^d) & T_\mr{uw}^{\mr{N}} & 0 \\ T_\mr{z y}^{\mr{N}} \mc{H}_L(y^d) & T_\mr{z w}^{\mr{N}} & -T_\mr{z}^{\mr{D}}} \Matrix{c}{g \\ w|_L \\ z|_L} = 0.
        \label{eq:LFRConstr}
    \end{equation}
    Additionally, the constraint from Lemma \ref{lem:init} is imposed and we note that \eqref{eq:LFRConstr} is expanded to
    \begin{equation}
        B \Matrix{c}{g \\ w|_L \\ z|_L} = 0,
    \end{equation}
    where $B$ is defined by \eqref{eq:B}. Then the LMI \eqref{eq:DDMBLMI} is obtained by using Finsler's Lemma.
\end{proof}
The LMI condition \eqref{eq:DDMBLMI} for verifying closed-loop dissipativity of arbitrary interconnections between data-driven and model-based representations will be used in the following section to verify the time-domain performance of controllers. 
We note that in this paper there has been no restriction to SISO systems, and the derived results thus apply to the MIMO case as well. Moreover, the model-based polynomial input-output representations describe the behavior exactly and no approximations are introduced. These novelties are in clear contrast to the results derived in \cite{wieler2021data}.

%% file: section/examples.tex
We now demonstrate the applicability of the result on two simulation examples: a SISO example with an integrator and a two-block mixed-sensitivity example. The first example shows that Theorem \ref{thm:DDMBLMI} can be used to approximate the $\mc{H}_{\infty}$-norm while the second example shows how time-domain performance of given controllers can be verified.
\subsection{SISO example}
In this example, we consider a SISO reference tracking example as shown in Fig.~\ref{fig:ReferenceTracking} with $W_\mr{S}=1, W_\mr{T}=0$, and as output the tracking error $e$. We consider a plant $\mc{G}$ and controller $\mc{K}$ with transfer functions
\begin{equation}
    G(q) = \frac{q+0.5}{q-0.5}, \quad K(q) = \frac{q+0.3}{q-1}.
\end{equation}
From this model-based representation of $\mc{G}$ we collect 300 input-output data points such that \eqref{eq:fundlemrank} is satisfied with horizon $L$. We extract $\mc{G}$ from the control configuration such that we consider the interconnection as the LFR structure shown in Fig.~\ref{fig:DDMBGenPlant}. Then we have that $\mc{M}$ has a rational representation defined as
\begin{equation}
    \Matrix{c}{u \\ z} = M(q) \Matrix{c}{y \\ w} = \Matrix{cc}{-K(q) & K(q) \\ -1 & 1}.
\end{equation}
Since we consider a SISO low-dimensional system, it follows that the input-output representation is defined by the polynomial matrices
\vspace{-5pt}
\begin{multline}
    \Matrix{c:c:c:c}{D_\mr{u}^\mr{m}(q) & 0 & N_\mr{u y}^\mr{m}(q) & N_\mr{u w}^\mr{m}(q) \\ \hdashline 0 & D_\mr{z}^\mr{m}(q) & N_\mr{z y}^\mr{m}(q) & N_\mr{z w}^\mr{m}(q)} = \\
    \Matrix{c:cc:c:c}{q - 1 & 0 & 0 & -q -0.3 & q+0.3\\ \hdashline 0 & q & 0 & -q & q\\ 0 & 0 & q & q & 0},
\end{multline}
which are indeed pair-wise coprime. Using model-based analysis on the interconnection between $\mc{G}$ and $\mc{M}$, we compute that the $\mc{H}_\infty$-norm is approximately equal to 1.0428. The finite-horizon $\ell_2$-gain is obtained by means of the LMI stated in Theorem~\ref{thm:DDMBLMI} by choosing $\nu = 3$ and letting the depth $L$ range from 3 to 40. In the top plot in Fig.~\ref{fig:FinL2Gain} we show that the finite-horizon $\ell_2$-gain converges for increasing $L$ to the $\mc{H}_{\infty}$-norm. This is compared to the model-based finite-horizon $\ell_2$-gain and note that both align. The derivation for the analysis of the model-based finite-horizon $\ell_2$-gain is described in the Appendix. It is to be expected that the finite-horizon $\ell_2$-gain is smaller than the $\mc{H}_{\infty}$-norm since finite trajectories can not completely characterize the worst-case behavior. \vspace{-5pt}
\subsection{Two-block mixed-sensitivity example}
\begin{figure}[t]
\centering
\includegraphics[width=\linewidth]{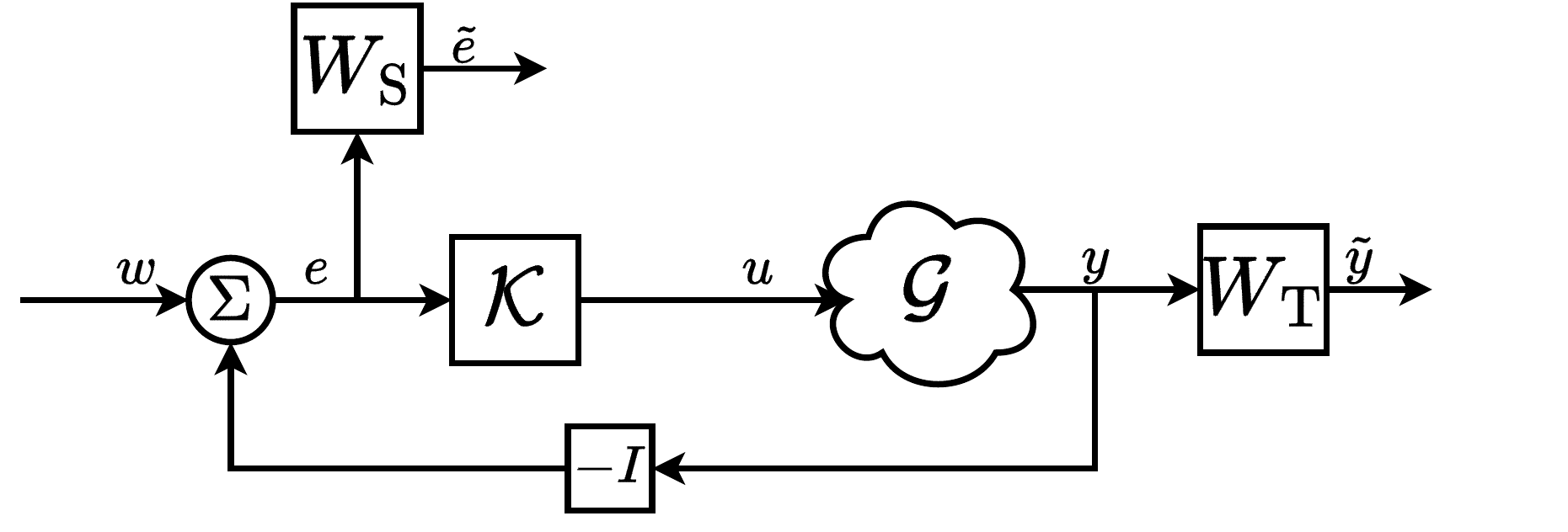} \vspace{-20pt}
\caption{Two-block reference tracking control configuration.} \vspace{-15pt}
\label{fig:ReferenceTracking}
\end{figure}
\begin{figure}[t]
\centering
\includegraphics[width=\linewidth]{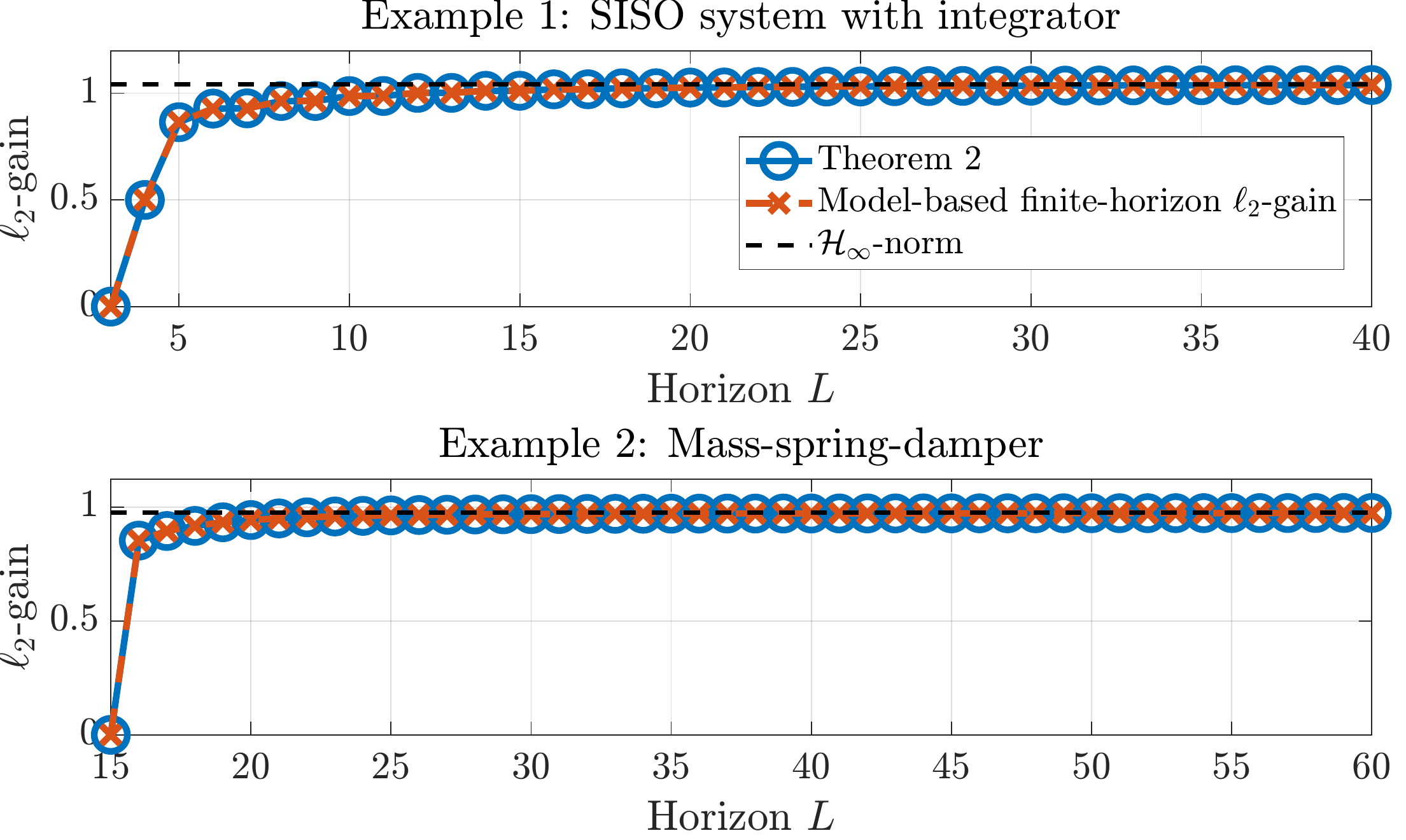}
\vspace{-20pt}\caption{Finite-horizon $\ell_2$-gain computed with the data-driven approach of Theorem \ref{thm:DDMBLMI} compared to the model-based finite-horizon $\ell_2$-gain.} \vspace{-10pt}
\label{fig:FinL2Gain}
\end{figure}

In this example, we consider SISO reference tracking as shown in Fig.~\ref{fig:ReferenceTracking} with the tracking error $e$ and plant output $y$ as outputs to shape the sensitivity and complementary sensitivity of the closed-loop behavior. We consider a two-mass-spring-damper system for $\mc{G}$ with state-space realization
\vspace{-15pt}
\begin{equation}
\begin{aligned}
    \dot{x} &= \Matrix{cccc}{0 & 1 & 0 & 0 \\ 
    -\frac{k_1+k_2}{m_1} & -\frac{d_1+d_2}{m_1} & \frac{k_2}{m_1} & \frac{d_2}{m_1} \\
    0 & 0 & 0 & 1 \\
    \frac{k_2}{m_2} & \frac{d_2}{m_2} & -\frac{k_2}{m_2} & -\frac{d_2}{m_2}}x + \Matrix{c}{0 \\ 0 \\ 0 \\ \frac{1}{m_2}} u,\\
    y &= \Matrix{cccc}{1 & 0 & 0 & 0} x,
\end{aligned}
\end{equation}
which is discretized with a sampling time $h=0.1 s$. The parameters have the following values: $m_1 = 10$ kg, $m_2 = 0.5$ kg, $d_1 = 200$ Ns/m, $d_2 =10$ Ns/m, $k_1 = 3000$ N/m, and $k_2 = 1000$ N/m. The generalized plant $\mc{P}$ is defined by the rational {matrix} 
\begin{equation}
    P(q) = \Matrix{cc}{W_\mr{S}(q) & -W_\mr{S}(q) G(q) \\ 0 & W_\mr{T}(q) G(q) \\ 1 & - G(q)},
\end{equation}
which is used to compute an $\mc{H}_{\infty}${-optimal} controller $K(q)$. The weighting filters $W_\mr{S}(q)$ and $W_\mr{T}(q)$ are chosen as
\begin{equation}
\resizebox{.85\hsize}{!}{$
     W_\mr{S}(q) = \frac{0.7741 q - 0.7641}{q-0.9998}, \quad W_\mr{T}(q) = \frac{25.9 q - 25.38}{q-0.3333},$}
\end{equation}
which specify a rise-time around 1 [s] and a bandwidth around 1.68 [rad/s]. The resulting $\mc{H}_{\infty}$-norm, computed with model-based analysis, is approximately equal to 0.9779. Next, we collect 400 input-output data points of $G(q)$ and extract the data-based representation of $\mc{G}$ into the upper block shown in Figure~\ref{fig:DDMBGenPlant}. Subsequently, $\mc{M}$ has rational {matrix}
\begin{equation}
    M(q) = \Matrix{cc}{-K(q) & K(q) \\ -W_\mr{S}(q) & W_\mr{S}(q) \\ W_\mr{T}(q) & 0}.
\end{equation}
The rational {matrix} 
representation $M(q)$ is converted to an input-output representation with the algorithm presented in \cite{antsaklis2007linear}. The finite-horizon $\ell_2$-gain calculated with Theorem~\ref{thm:DDMBLMI} is compared to the model-based gain for $\nu=15$ and a depth $L$ ranging from $\nu$ to 60. This comparison is shown in the bottom plot of Fig.~\ref{fig:FinL2Gain}, which shows that both computed finite-horizon $\ell_2$-gains are equivalent and converge to the $\mc{H}_{\infty}$-norm.
\vspace{-6pt}

%% file: section/conclusion.tex
\vspace{-1pt}
In this paper, we have generalized interconnections between data-driven and model-based representations in an LFR form on a finite horizon. This LFR structure is specifically leveraged to define arbitrary MIMO interconnection between data-driven representations of systems, for which only measured data is available, and model-based representations of controllers and weighting filters. Moreover, tractable LMI-based dissipativity analysis results have been derived for these LFR-based combined representations. Through mixed-sensitivity arguments, time-domain performance of designed controllers, for data-driven representations, can be verified a priori. For future work, we aim to extend this framework towards controller synthesis with guaranteed performance.

%% file: section/appendix.tex
Given a minimal state-space realization \eqref{eq:sys} of an LTI system $\mc{G}$, define the finite-horizon observability matrix and the Toeplitz matrix as
\vspace{-10pt}
\begin{equation}
\mc{O}_L(\mc{G}) = \begin{bsmallmatrix}
C \\ C A \\ \svdots \\ C A^{L-1}
\end{bsmallmatrix},
    \mc{T}_L(\mc{G}) = \begin{bsmallmatrix}
    D & 0 & \cdots & 0 \\ C B & D & \sddots & \svdots \\ \svdots & \sddots & \sddots & 0\\ 
    C A^{L-2} B & \cdots & C B & D
    \end{bsmallmatrix} .\vspace{-5pt}
\end{equation}
The finite-horizon input-output behavior of the system is defined by
\vspace{-5pt}
\begin{equation}
    y|_L = \mc{O}_L(\mc{G}) x_0 + \mc{T}_L(\mc{G}) u|_L.\vspace{-5pt}
\end{equation}
If a zero initial condition, e.g., $x_0 = 0$, is assumed we have that
\vspace{-5pt}
\begin{equation}
    \Matrix{c}{u|_L \\ y|_L} = \Matrix{c}{I \\ \mc{T}_L(\mc{G})} u|_L. \vspace{-5pt}
\end{equation}
The dissipativity inequality then becomes
\vspace{-5pt}
\begin{equation}
    (*)^{\top} \Pi_L \Matrix{c}{I \\ \mc{T}_L(\mc{G})} u|_L \geq 0, \vspace{-5pt}
\end{equation}
and we conclude that model-based $L$-dissipativity can be verified by
\vspace{-10pt}
\begin{equation}
    \Matrix{c}{I \\ \mc{T}_L(\mc{G})}^{\top} \Pi_L \Matrix{c}{I \\ \mc{T}_L(\mc{G})} \succeq 0.
\end{equation}